\documentclass[conference]{IEEEtran}
\usepackage{amsmath}
\usepackage{amssymb}
\usepackage{amsfonts}
\usepackage{amsthm}
\usepackage{algorithm}
\usepackage{algorithmic}
\usepackage{graphicx}
\usepackage{array}

\usepackage[shortlabels]{enumitem}
\usepackage{cite}
\usepackage{color}
\usepackage{epsfig,epstopdf}
\usepackage{url}
\usepackage{subfigure}
\usepackage{balance}
\usepackage{multirow}
\usepackage{verbatim} 
\usepackage{makecell}

\IEEEoverridecommandlockouts

\begin{document}

\title{Sensor Deployment and Link Analysis in Satellite IoT Systems for Wildfire Detection}

\author{\IEEEauthorblockN{How-Hang Liu{\textsuperscript{$1$}}, Ronald Y. Chang{\textsuperscript{$1,2$}}, Yi-Ying Chen{\textsuperscript{$3$}}, I-Kang Fu{\textsuperscript{$4$}}, and H. Vincent Poor{\textsuperscript{$2$}}}
\IEEEauthorblockA{
{\textsuperscript{$1$}}Research Center for Information Technology Innovation, Academia Sinica, Taiwan \\
{\textsuperscript{$2$}}Department of Electrical and Computer Engineering, Princeton University, USA \\
{\textsuperscript{$3$}}Research Center for Environmental Changes, Academia Sinica, Taiwan \\
{\textsuperscript{$4$}}MediaTek Inc. \\
}
\IEEEauthorblockA{Email: \{liuhowhang, rchang\}@citi.sinica.edu.tw, yiyingchen@gate.sinica.edu.tw, ik.fu@mediatek.com, poor@princeton.edu}
\thanks{This work was supported in part by the Ministry of Science and Technology, Taiwan, under Grants MOST 109-2221-E-001-013-MY3 and MOST 109-2918-I-001-003.}}

\maketitle

\begin{abstract}
Climate change has been identified as one of the most critical threats to human civilization and sustainability. Wildfires, which produce huge amounts of carbon emission, are both drivers and results of climate change. An early and timely wildfire detection system can constrain fires to short and small ones and yield significant carbon reduction. In this paper, we propose to use ground sensor deployment and satellite Internet of Things (IoT) technologies for wildfire detection by taking advantage of satellites' ubiquitous global coverage. We first develop an optimal IoT sensor placement strategy based on fire ignition and detection models. Then, we analyze the uplink satellite communication budget and the bandwidth required for wildfire detection under the narrowband IoT (NB-IoT) radio interface. Finally, we conduct simulations on the California wildfire database and quantify the potential economical benefits by factoring in carbon emission reductions and sensor/bandwidth costs. 
\end{abstract}


\IEEEpeerreviewmaketitle

\section{Introduction}\label{sec:intro}

Non-terrestrial networks (NTNs) are envisioned to complement terrestrial networks in providing connection services over remote or infrastructure-deficient areas in the era of Internet of Things (IoT). NTNs can generally be classified as low Earth orbit (LEO) and geostationary orbit (GEO) systems with different altitudes and beam sizes of the satellites \cite{NTN5G}. GEO satellites' orbital period is synchronized with Earth's rotational period with an altitude and spot beam diameter being $35786$ km and $100$--$3500$ km, respectively. LEO satellites have an altitude of $500$--$2000$ km and spot beam diameter of $50$--$1000$ km.

The 3rd generation partnership project (3GPP) has studied the evolution of fifth-generation (5G) wireless technologies to support NTN \cite{3gppNTN}. In particular, 3GPP Release 17 has studied the adaptation of narrowband IoT (NB-IoT), among other technologies, to support NTN. The features of NB-IoT and the challenges of applying NB-IoT to NTN were presented in \cite{NIOT_NTN}. The main challenges for adapting NB-IoT to GEO NTN include the large round-trip time (RTT) and path loss, and the main challenges for adapting NB-IoT to LEO NTN include Doppler shifts due to the high satellite velocity. Some use cases of NB-IoT NTN have been discussed \cite{22822}, including providing services for disaster response and relief operations in the event of outage or disruption of terrestrial networks.

A potential key application of NB-IoT NTN is wildfire detection. Wildfires are devastating disasters threatening lives, properties, and natural resources, and cause a huge amount of carbon emission which contributes to a nontrivial percentage of global carbon footprints. Traditional wildfire detection methods include satellite imaging, vision-based remote sensing, and wireless sensor network-based environment monitoring \cite{UAV_fire}. In \cite{CNN}, a vision-based detection method enhanced by convolutional neural networks (CNNs) was proposed. In \cite{UAV_fire}, an unmanned aerial vehicle assisted IoT (UAV-IoT) network was proposed to complement the satellite imaging technology for wildfire detection, where the wildfire detection probability was analyzed based on geometry and probability theory. In our previous work \cite{sensorbasedIOT}, a sensor-based NB-IoT NTN system for wildfire detection was proposed, where a wildfire evolution model was developed and monetary savings as a result of sensor-based early wildfire detection were calculated. However, sensor placement strategies and communication bandwidth costs were not considered.

In this work, we consider sensor-based NB-IoT NTN with a GEO satellite for wildfire detection, and investigate specifically sensor placement and satellite IoT system capacity. We develop an optimal sensor placement strategy that takes into account fire ignition probability (related to biomass, soil moisture, and lightning/human causes) and fire detection probability (related to sensor density). We perform a satellite communication budget analysis by deriving the uplink signal-to-noise ratio (SNR) for IoT devices at different beam locations, and deriving the amount of bandwidth required for simultaneous data transmission among a number of IoT devices according to an IoT traffic model. Finally, we conduct an independent simulation based on historical California wildfires, which demonstrates results that have direct implications on greenhouse gas emission reduction and economical benefits. The proposed method can be facilitated by the proven feasibility of using consumer-grade devices for NTN connection \cite{SpaceIoT}, and can guide new objectives and applications in NTN technology development.

\section{Wildfire Model and Optimal Sensor Placement} \label{sec:SM}

\begin{figure}[t]
\centering
\includegraphics[width=0.77\columnwidth]{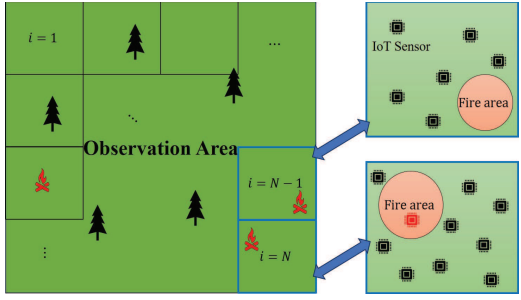}
\caption{Sensor deployment over a wide area with $N$ equal-area local regions for wildfire detection. Fire detection is modeled as the event that at least one sensor falls inside the fire area. The events of detecting (bottom right) and not detecting (top right) fires are illustrated.}
\label{fig:systemModel}
\vspace{-0.1in}
\end{figure}

We consider a total of $K$ sensors being deployed over a wide area with $N$ equal-area local regions, as shown in Fig.~\ref{fig:systemModel}. Each region $i$ is deployed with $n_i$ sensors. Each region is characterized by some environmental factors that collectively contribute to a projected fire ignition probability of the region. The objective of sensor placement is to deploy a proper number of sensors in each region by taking into account the fire ignition probability of the region, such that some overall utility related to fire detection in the entire area is maximized. 
The fire ignition probability $p_i^{I}$ of region $i$ depends on the biomass, soil moisture, and lightning/human causes in the region, and can be expressed as \cite{FireSpread}
\begin{equation} \label{eq:p_FI}
p_i^{I}=p_i^B \times p_i^M \times p_i^L.
\end{equation}
The $p_i^B$ is the probability of fire depending on the available biomass, which is given by
\begin{equation} \label{eq:p_b}
p_i^B = \max\left[ 0, \min\left (1, \frac{B_i-B_{\rm low}}{B_{\rm up}-B_{\rm low}} \right ) \right]
\end{equation}
where $B_i$ is the above-ground biomass (KgC/m$^{\text{2}}$) of region $i$, $B_{\rm low}=0.2$ KgC/m$^{\text{2}}$ is the lower biomass threshold (below which fire will not occur, i.e., $p_i^B=0$), and $B_{\rm up}=1$ KgC/m$^{\text{2}}$ is the upper biomass threshold (above which $p_i^B=1$). The $p_i^M$ is the probability of fire depending on the soil moisture, expressed in terms of the root zone soil wetness $\beta_{{\rm root},i}=\max\left[ 0, \min\left(1,\frac{\theta_{i}-\theta_{\rm wilt}}{\theta_{\rm field}-\theta_{\rm wilt}} \right ) \right]$, where $\theta_{i}$ is the volumetric soil moisture content, and $\theta_{\rm wilt}$ and $\theta_{\rm field}$ represent wilting point and field capacity soil moisture contents, respectively. The dependence of $p_i^M$ on $\beta_{{\rm root},i}$ is represented as 
\begin{equation} \label{eq:p_w}
p_i^M=1-\tanh\left(\frac{1.75\beta_{{\rm root},i}}{\beta_e}\right)^2
\end{equation}
where $\beta_e=0.35$ is an extinction wetness content. The $p_i^L$ is the probability of fire due to lightning/human causes. First, define a lightning scalar similar to \eqref{eq:p_b} as $\beta_{L,i}=\max\left[0,\min\left(1,\frac{L_{i}-L_{\rm low}}{L_{\rm up}-L_{\rm low}} \right) \right]$, where $L_i$ is the cloud-to-ground lightning frequency (flashes/km$^{\text{2}}$/month) of region $i$, $L_{\rm low}=0.02$ flashes/km$^{\text{2}}$/month, and $L_{\rm up}=0.85$ flashes/km$^{\text{2}}$/month. Then, the dependence of $p_i^L$ on $\beta_{L,i}$ is expressed as
\begin{equation} \label{eq:I}
p_i^L = I_i + (1-I_i) p_i^H
\end{equation}
where $I_i = \frac{\beta_{L,i}}{\beta_{L,i}+\exp(1.5-6\beta_{L,i})}$ and $p_i^H$ is the probability of fire ignition due to human causes. Note that $I_i\in [0,1]$, and in the absence of lightning (i.e., $\beta_{L,i}=0$ and thus $I_i=0$), $p_i^L=p_i^H$.

For simplicity, we define fire detection as the event that at least one sensor falls inside the fire area, as shown in Fig.~\ref{fig:systemModel}. We assume each fire occurs independently. Let $p_i^{D}(t)$ be the probability of fire detection in region $i$ at time $t$, where $t$ is the duration (in the unit of hours) from fire ignition to fire detection. The probability of fire detection in region $i$ at $t=T$ is computed by
\begin{equation} \label{eq:p_FD}
p_i^{D}(T)=1-\left(\frac{\max\big(0, A-a_i(T)\big)}{A}\right)^{n_i}
\end{equation}
where $A$ is the area of each region and $a_i(T)$ is the fire burned area in region $i$ at time $T$. The second term of \eqref{eq:p_FD} represents the probability that no sensor (out of $n_i$ sensors) is inside the fire area in region $i$ at time $T$. The fire burned area is modeled by a circle, i.e., $a_i(T)=\pi(u_{p,i}T)^2$, where $u_{p,i}$ is the average fire spreading rate in region $i$ and $u_{p,i}T$ is the radius of the circular fire area. 

Define the overall {\it system utility} at some specific time $T$ as
\begin{equation} \label{eq:p_det}
U(T) = \sum_{i=1}^{N}p_i^{I}\times p_i^{D}(T).
\end{equation}
The sensor placement design problem is
\begin{subequations} \label{eq:opt}
\begin{align}
\max_{\{n_i\}} & & & U(T) \label{eq:opt_obj} \\
\text{s.t.} & & & n_i\in\mathbb{Z}=\{0,1,2,\ldots\}, \forall i \label{eq:opt_cons1} \\
& & & \sum_{i=1}^{N} n_i \leq K \label{eq:opt_cons2}
\end{align}
\end{subequations}
where constraints \eqref{eq:opt_cons1} and \eqref{eq:opt_cons2} specify the integer number of sensors in each region and the total number of sensors, respectively. A solution that aligns a higher fire detection probability with a higher projected fire ignition probability in all regions to maximize the system utility is expected.

\section{Satellite Communication Budget Analysis for Wildfire Detection} \label{sec:SCM}

\subsection{Satellite Communication Model}

\begin{figure}[t]
\centering
\includegraphics[width=0.92\columnwidth]{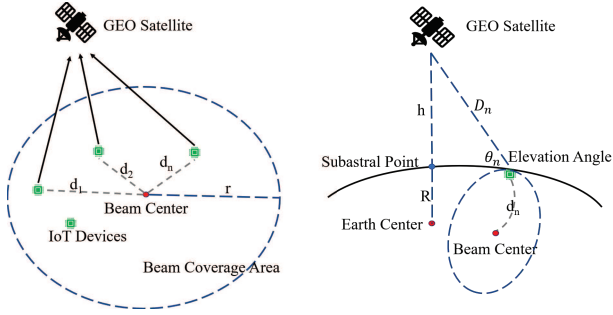}
\caption{GEO satellite communication system serving wildfire-detecting IoT sensor devices.}
\label{fig:satelliteModel}
\vspace{-0.1in}
\end{figure}

We consider a single-beam GEO satellite communication system serving IoT sensor devices, as shown in Fig.~\ref{fig:satelliteModel}. The spot beam coverage area is modeled as a circle area with a beam center and beam radius $r$. The distances of IoT devices to the beam center are denoted as $d_n$'s. The $h$, $D_n$, $R$, and $\theta_n$ in Fig.~\ref{fig:satelliteModel} denote the satellite height, the distance between IoT device $n$ and the satellite, radius of earth, and the elevation angle of IoT device $n$, respectively. Note that all distances are in the unit of km and all angles are in the unit of degree. The IoT devices in the spot beam use frequency division multiple access (FDMA) to avoid the intra-beam interference. Radio access is based on 3GPP NB-IoT with a $180$ kHz bandwidth, using orthogonal frequency-division multiple access (OFDMA) for downlink and single-carrier frequency-division multiple access (SC-FDMA) for uplink.
The transmitted symbol from IoT device $n$ to the satellite on subcarrier $k$ is denoted by $x_{n,k}\in \mathbb{C}^{1\times 1}$ with unit power. The received signal at the satellite on subcarrier $k$ can be expressed as \cite{PGR}
\begin{equation} \label{eq:R_signal}
y=\sqrt{P_{n,k}G_{n,k}}x_{n,k}+n_0
\end{equation}
where $P_{n,k}$ is the transmit power of IoT device $n$ on subcarrier $k$, $G_{n,k}$ is the channel gain, and $n_0 \in {\cal CN}(0,\sigma^2)$ is the additive white Gaussian noise (AWGN). The channel gain includes the antenna gain, beam gain, free-space path loss and other losses, and small-scale fading, i.e., 
\begin{equation} \label{eq:channelGain}
G_{n,k}=g_n(\varepsilon) G(d_n) L_n f_\varepsilon \left|h_{n,k}\right|^2
\end{equation}
where $g_n(\varepsilon)$ is the antenna gain that depends on the off-boresight angle $\varepsilon$ from the IoT device to the satellite, i.e.,
\begin{equation} \label{eq:antennaGain}
g_n(\varepsilon)~[\rm dBi]=\begin{cases}
G_{t,\rm max}, & 0^\circ<\varepsilon\leq1^\circ \\
32-25\log\varepsilon, & 1^\circ <\varepsilon\leq48^\circ \\
-10, & 48^\circ<\varepsilon\leq180^\circ \\
\end{cases}
\end{equation}
with $G_{t,\rm max}$ being the maximal terrestrial antenna gain of main lobe. The $G(d_n)$ is the beam gain modeled as \cite{ECN}
\begin{equation} \label{eq:beamGain}
G(d_n)~[{\rm dBi}]=G_{s,\rm max}\begin{pmatrix}
\frac{J_1(ad_n)}{2ad_n}+36\frac{J_3(ad_n)}{a^3d_n^3} 
\end{pmatrix}^2
\end{equation}
where $G_{s,\max}$ is the maximum antenna gain of the satellite, $J_m(\cdot)$ is the Bessel function of the first kind of order $m$, $a=2.07123/r$ (where $r$ is the beam radius). The $L_n=(c/(4\pi f D_n))^2$ is the free-space path loss for IoT device $n$, with $c$, $f$, and $D_n$ being the light speed, carrier frequency, and straight-line distance from the IoT device $n$ to the satellite. The $f_\varepsilon$ represents other losses including atmospheric path loss, scintillation loss, and polarization loss. The $h_{n,k}$ is the small-scale fading assumed to follow the Shadowed-Rician fading model \cite{ANS}, with the probability density function (PDF) of $|h_{n,k}|^2$ for all $k$ given by
\begin{equation} \label{eq:fading}
f_{|h_{n,k}|^2}(x)=\alpha {\rm exp}(-\beta x)_1F_1(m,1,\delta x)
\end{equation}
where $_1F_1(\cdot,\cdot,\cdot)$ denotes the confluent hypergeometric function, $\alpha=\frac{(2bm)^m}{2b(2bm+\zeta)^m}$, $\delta=\frac{\zeta }{2b(2bm+\zeta)}$, and $\beta=\frac{1}{2b}$, with $2b$ being the average power of the scatter component, $m$ denoting the Nakagami fading parameter, and $\zeta$ denoting the average power of the line-of-sight (LoS) component. Note that parameters $b$, $m$, and $\zeta$ depend on the elevation angle $\theta_n$, as explicitly specified by \cite{ANS}
\begin{equation} \label{eq:parameterwithTheta}
\begin{split}
b(\theta_n)&=-4.7943\times 10^{-8}\theta_n^3+5.5784\times10^{-6}\theta_n^2 \\
&\quad -2.1344 \times 10^{-4}\theta_n+3.271\times10^{-2}, \\
m(\theta_n)&=6.3739\times 10^{-5}\theta_n^3+5.8533\times10^{-4}\theta_n^2 \\
&\quad -1.5973 \times 10^{-1}\theta_n+3.5156, \\
\zeta(\theta_n)&=1.4428\times 10^{-5}\theta_n^3-2.3798\times 10^{-3}\theta_n^2 \\
&\quad +1.2702 \times 10^{-1}\theta_n-1.4864. \\
\end{split}
\end{equation}

\subsection{IoT Traffic Model}\label{Traffic}

We consider two types of traffic models for IoT sensor devices \cite[Secs. E.2.1 and E.2.2]{45820}, namely, {\it exception reports} for emergency event reporting (e.g., a wildfire outburst), and {\it periodic reports} for regular monitoring. For exception reports, the uplink application payload is $20$ bytes per sensor report, and such reports are required near real-time with a latency target of $10$ seconds. Considering $10^6$ sensors simultaneously sending exception reports, the total amount of data transmitted is $20\times 10^6=2\times 10^7$ bytes in $10$ seconds. For periodic reports, the number of transmission sessions is first calculated: $S=11.2\times K\times (T_{\rm obs}/86400)$, where $K$ is the number of sensors and $T_{\rm obs}$ (in second) is the observation time in the fraction of a day. The payload of periodic reports ranges from $20$ to $200$ bytes. For comparison, we assume $10$ seconds as the reference period and $20$ bytes as the payload size for both types of traffic. Considering $10^6$ sensors and $T_{\rm obs}=10$, we have $S\approx 1296$ and for each session $20$ bytes are transmitted. Thus, the total amount of data transmitted is about $1296\times 20=2.592\times 10^4$ bytes in $10$ seconds.

\subsection{SNR Analysis}\label{SNR}

The uplink SNR for IoT device $n$ on subcarrier $k$ is ${\rm SNR}_{n,k}=\frac{P_{n,k}G_{n,k}}{\sigma^2}$. We consider IoT sensor devices located at the edge and center of the spot beam with the largest and smallest distances to beam center $d_n$, respectively, for the worst- and best-case SNR analysis, since $G(d_n)$ is nonincreasing in $d_n$ and it dominates $G_{n,k}$ and thus the SNR. We consider the Galaxy 30 satellite \cite{galaxy} located at $125^\circ \rm{W}$ with the beam center at ($37^\circ \rm{N}, 122^\circ \rm{W}$), and the edge and center IoT sensor devices located at ($33.5^\circ \rm{N}, 116.6^\circ \rm{W}$) and ($37.2^\circ \rm{N}, 122.1^\circ \rm{W}$), respectively. The corresponding parameters are listed in Table~\ref{tab:simPar}. The SNR can be calculated as $-0.45$ dB and $5.55$ dB for edge and center IoT devices, which correspond to modulation and coding scheme (MCS) levels 5 and 11, respectively \cite{NBIOT_GEO}. The higher the MCS level, the more bits can be transmitted in a resource unit (RU).

	\begin{table}[t]
	\begin{center}
	\caption{Satellite Communication Parameters}
	\label{tab:simPar} \vspace*{0in}
	\setlength{\tabcolsep}{1mm}{
	\begin{tabular}{|c||c|c|}
	\hline
	Parameters & Edge IoT & Center IoT \\ \hline\hline
	Orbit & GEO & GEO \\ \hline
	Elevation angle ($\theta_n$) & $50^\circ$ & $46.8^\circ$\\ \hline
	Off-boresight angle ($\varepsilon$) & $50^\circ$ & $50^\circ$\\ \hline
	Distance to satellite ($D_n$)& $37123$ km & $37353$ km\\ \hline
	Distance to beam center ($d_n$)& $639$ km & $24$ km\\ \hline
    Carrier frequency ($f$)& $2$ GHz & $2$ GHz\\ \hline
	Max satellite antenna gain ($G_{s,{\rm max}}$) & $25$ dBi & $25$ dBi\\ \hline
	Max IoT device antenna gain ($G_{t,{\rm max}}$) & $7.38$ dBi & $7.38$ dBi \\ \hline
	Beam radius ($r$) & $1000$ km & $1000$ km\\ \hline
	$b$ & $0.03$ & $0.03$\\ \hline
	$m$ & $4.96$ & $3.86$\\ \hline
	$\zeta$ & $0.72$  & $0.72$\\ \hline
	Uplink power ($P_{n,k}$) & $23$ dBm & $23$ dBm \\ \hline
	Noise power ($\sigma^2$) & $-167.42$ dBm & $-167.42$ dBm\\ \hline
	Other losses ($f_\varepsilon$) & $-10$ dB & $-10$ dB\\ \hline
	SNR & $-0.45$ dB & $5.55$ dB\\ \hline
	\end{tabular}}
	\end{center}
	\vspace{-0.2in}
	\end{table}

\subsection{Bandwidth Requirements}\label{RA}

Here, we analyze the satellite communication bandwidth requirement. We consider the worst-case SNR ($-0.45$ dB) to obtain the worst-case (maximum) bandwidth requirement. To fulfill the uplink payload of $20$ bytes, each uplink transmission at MCS level 5 requires three RUs \cite[Table I]{NBLEO}. The uplink transmission process requires at least two RTTs if there is no retransmission \cite[Sec. 7.1.7.3]{45820}. The first RTT represents the duration of random access request and response between the IoT device and the satellite, and the second RTT represents the duration of uplink data transmission and acknowledgment between the IoT device and the satellite. Each RTT is assumed to be $500$ ms for GEO satellites \cite{NIOT_NTN}. Each RU for single-tone transmissions is $3.75$ kHz in frequency and $32$ ms in time \cite{NBIOT}. Thus, the total time requirement for a single IoT device to transmit a report is $500\times2+32\times3=1096$ ms.

Considering the reference period of $10$ seconds and NB-IoT bandwidth $180$ kHz, the supportable number of IoT sensor devices is $\lfloor 10000/1096\rfloor \times (180/3.75)=432$ for the exception report traffic. For the periodic report traffic, the number can be similarly calculated, which is $337824$ IoT sensor devices. Since the exception report traffic demands significantly more bandwidth per sensor as compared to the periodic report traffic, we consider the exception report traffic type for the worst-case bandwidth requirement. The results for two example systems are: a $10^5$-sensor system requires $\left \lceil(10^5/432)\right \rceil\times 180~(\mbox{kHz}) = 41.76$ MHz bandwidth, and a $10^6$-sensor system requires $\left \lceil(10^6/432)\right \rceil\times 180~(\mbox{kHz}) = 416.7$ MHz bandwidth. Since in obtaining these numbers we have assumed that {\it all IoT devices have the worst-case SNR and all transmit exception reports simultaneously}, the bandwidth requirement in practice could be much lower than these numbers.

The cost of satellite communication spectrum depends on the spectrum price setting and trading which is a complicated process. We follow the adoption of a monetary coefficient $0.6$ (USD/Hz) in \cite{STSC}, i.e., the cost of bandwidth in USD is $0.6$ times of the bandwidth in Hz. Thus, the bandwidth cost is $41.76~(\mbox{MHz})\times 0.6=25$ million USD for a $10^5$-sensor system, and $416.7~(\mbox{MHz})\times 0.6=250$ million USD for a $10^6$-sensor system.

\section{Simulation Results and Discussion}

\begin{figure}[t]
\centering
\includegraphics[width=0.85\columnwidth]{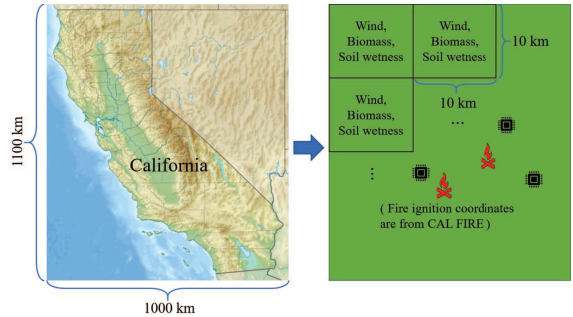}
\caption{Simulation settings.}
\label{fig:simulationSettings}
\vspace{-0.1in}
\end{figure}

\subsection{Simulation Settings} \label{sec:SS}

We simulate the proposed system using California wildfires as an exemplary study. Since California spans about $1100$ km latitudinally (vertically) and $1000$ km longitudinally (horizontally), we consider a $1100\times 1000$ km$^{\text{2}}$ geographic area and partition it into $10\times 10$ km$^{\text{2}}$ local regions, as shown in Fig.~\ref{fig:simulationSettings}. There are $N=11000$ regions, each of area $A=100$ km$^{\text{2}}$ and with associated wind, biomass, and soil wetness data \cite{CDS}. We set $p_i^H=0.5$ for each region and $T=4$ in \eqref{eq:opt} as the default target time for fire detection, as undetected fires beyond four hours could be uncontrollable \cite{DWSE}. We set the coordinates of fire ignition points according to the real fire ignition locations of $255$ wildfires in the 2020 annual records from the CAL FIRE database \cite{CALFIRE}, which are {\it independent} of the fire ignition probability model in Sec.~\ref{sec:SM}. The total number of sensors $K$ ranges from $10^5$ to $10^6$, which is approximately $0.24$ and $2.4$ sensors per km$^{\text{2}}$ if all sensors are placed in the geographic area of California.

We consider the following sensor placement schemes: 1) {\it biomass uniform placement}, where the total number of sensors are allocated equally to all regions for which biomass is greater than zero (i.e., $B_i > 0$ in \eqref{eq:p_b}); and 2) {\it optimized placement}, where the total number of sensors are allocated to the regions according to the solution to problem~\eqref{eq:opt}. CVX \cite{cvx} is applied to solve the problem. For each scheme, the sensors are uniformly distributed inside each region.

\subsection{Results and Discussion} \label{sec:SR}

\begin{figure*}[t]
\begin{center}
\subfigure[]{
    \label{fig:detection_utility}
    \includegraphics[width=0.85\columnwidth]{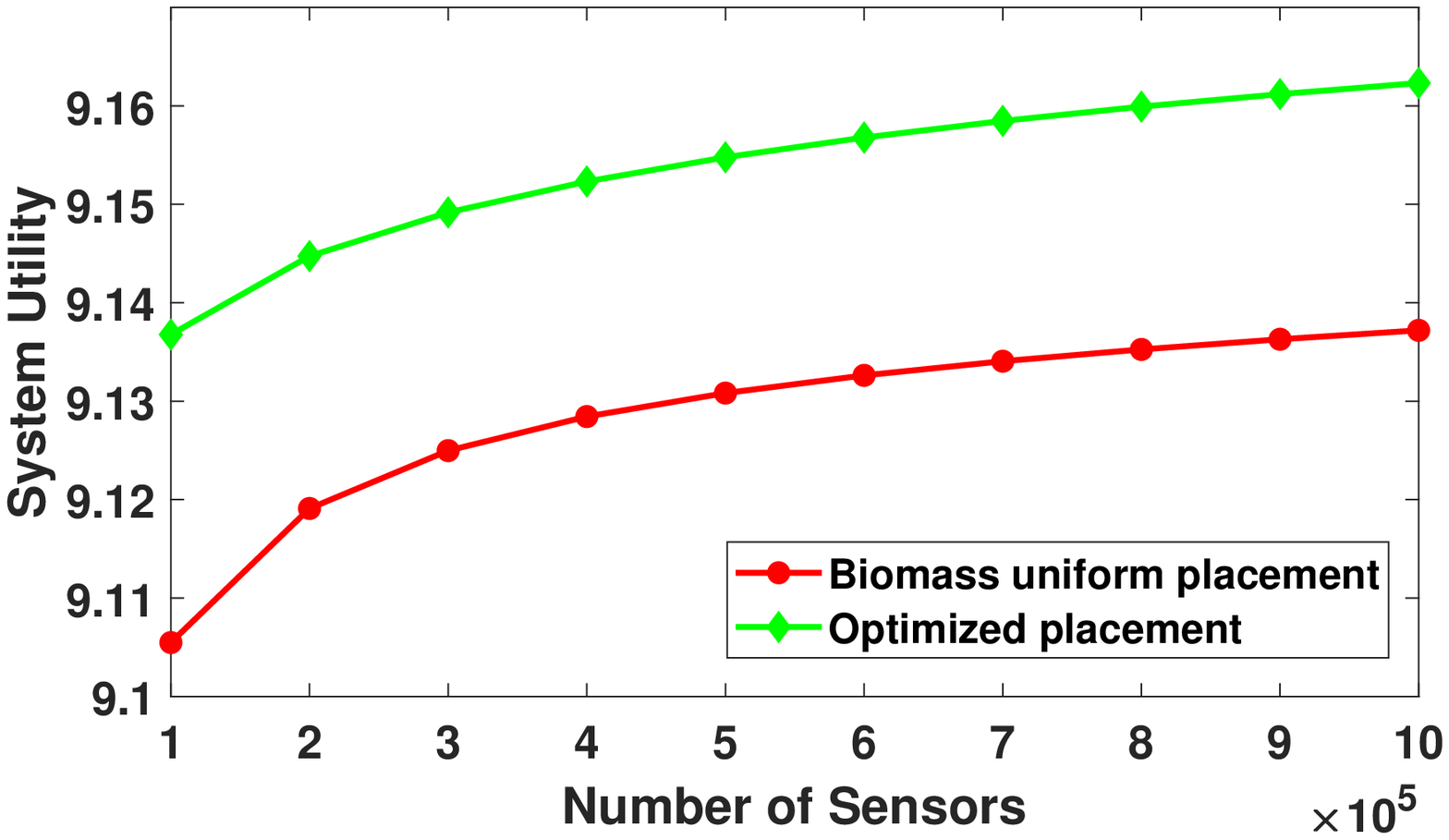}}
\subfigure[]{
    \label{fig:compare_area}
    \includegraphics[width=0.85\columnwidth]{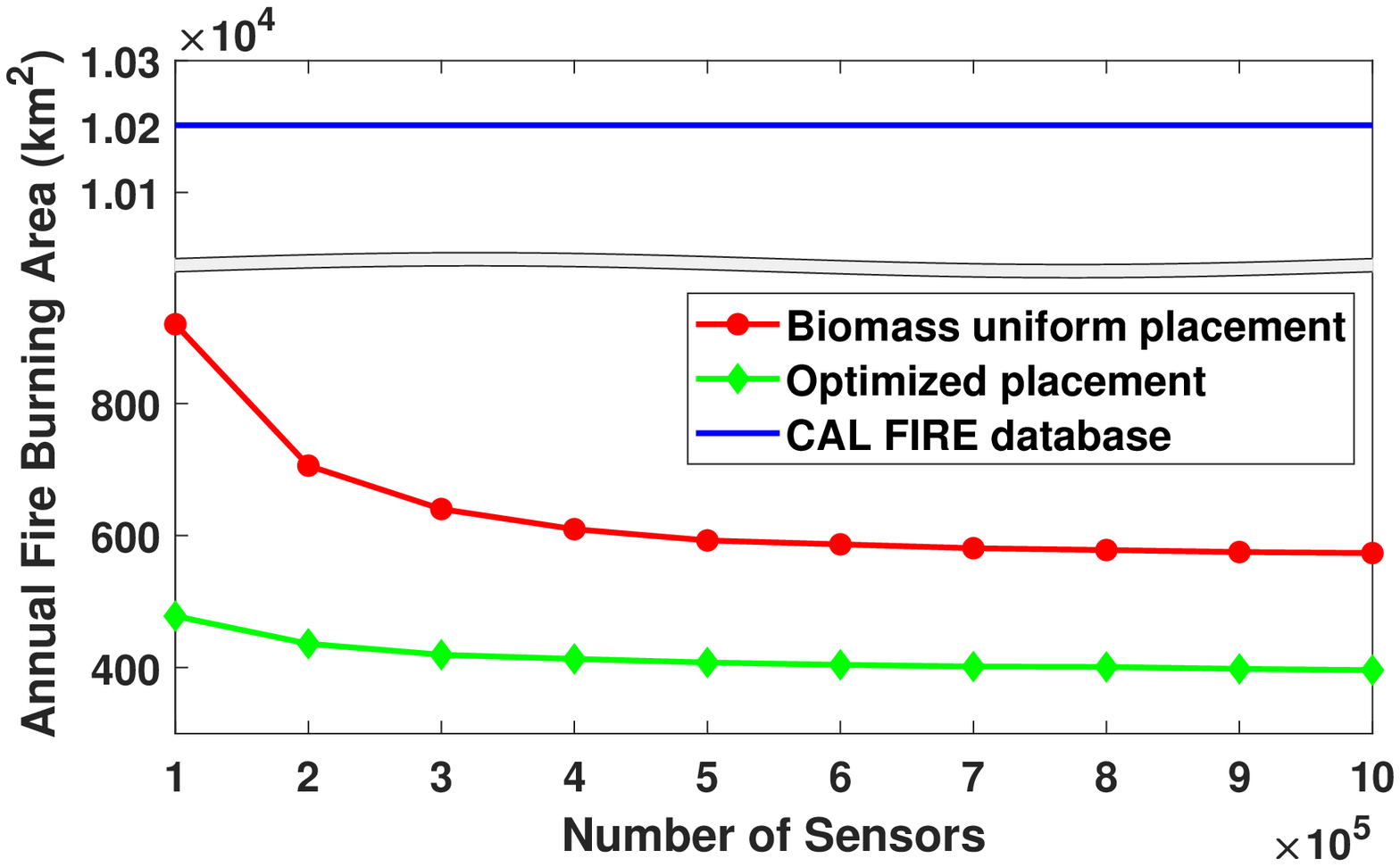}}
\subfigure[]{
    \label{fig:compare_carbon}
    \includegraphics[width=0.85\columnwidth]{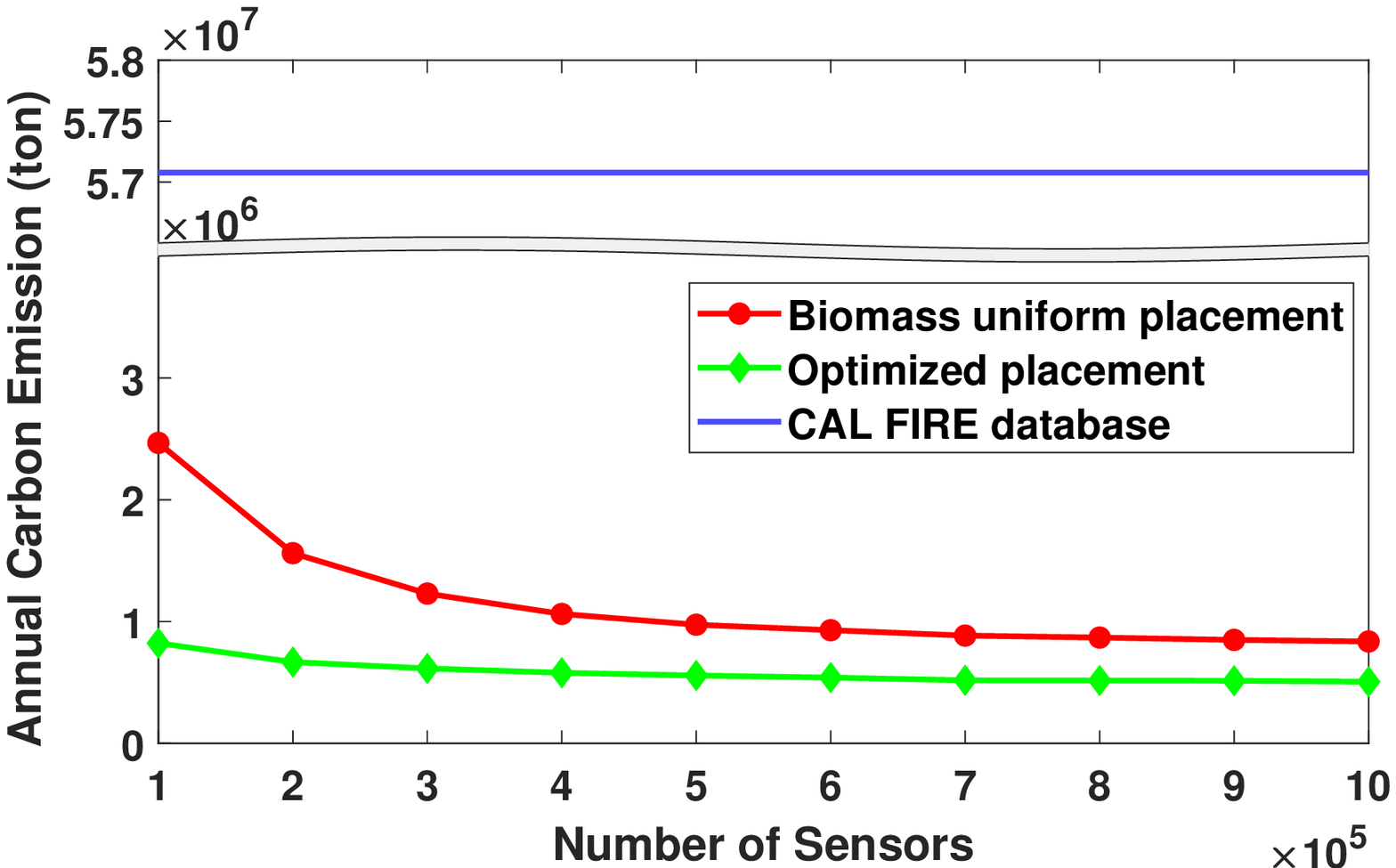}}
\subfigure[]{
    \label{fig:compare_savings}
    \includegraphics[width=0.85\columnwidth]{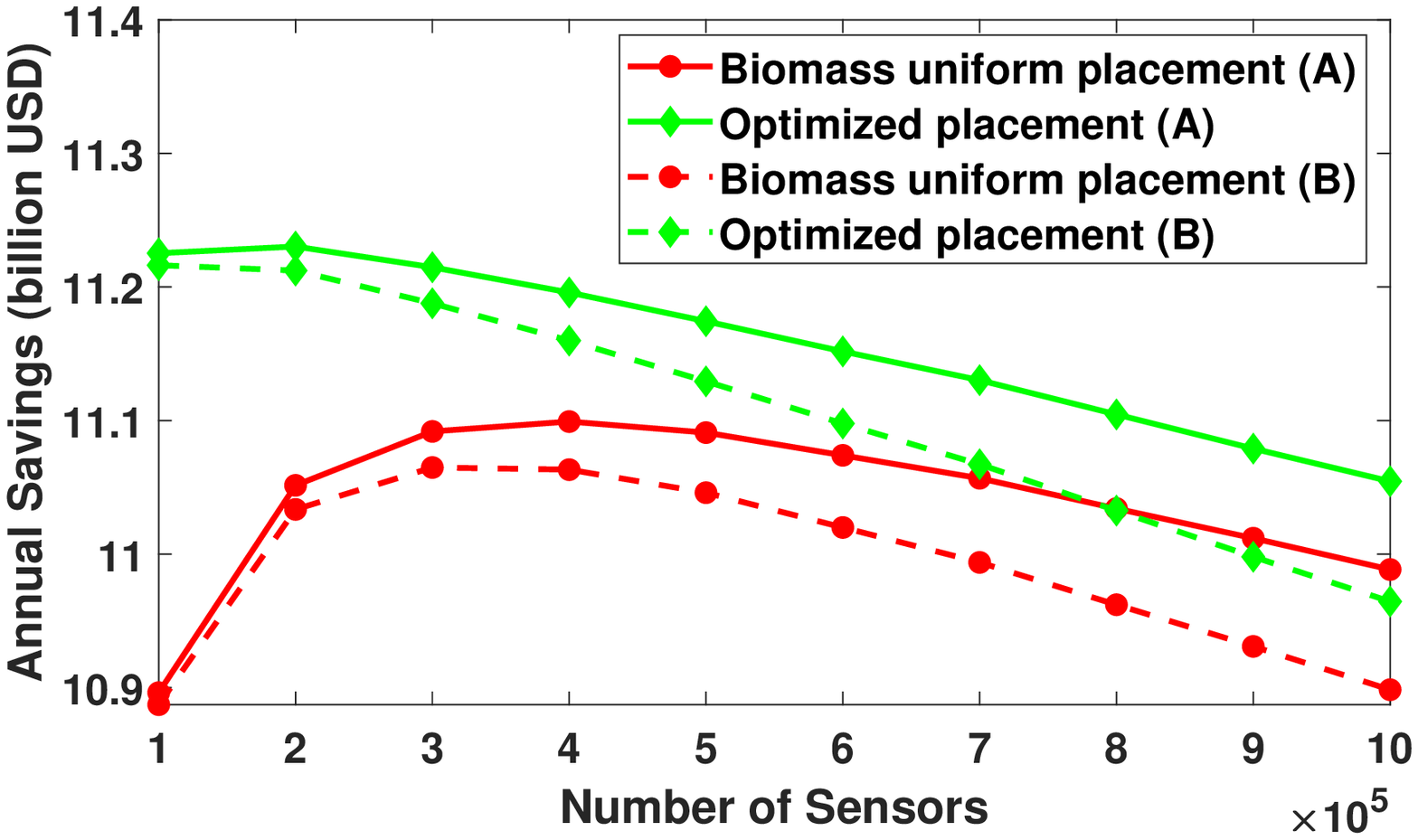}}
\caption{Simulation results. (a) System utility (in \eqref{eq:p_det}). (b) Annual fire burning area (km$^{\text{2}}$). (c) Annual carbon emission (ton). (d) Annual savings (billion USD) for device costs of $10$ USD (case A) and $100$ USD (case B) per device.}
\label{fig:simulation_result}
\end{center}
\vspace{-0.1in}
\end{figure*}

{\bf System utility.} Fig.~\ref{fig:detection_utility} compares the system utility in \eqref{eq:p_det} for the two placement schemes. As the sensor number increases, the system utility increases because the probability of fire detection $p_i^{D}(T)$ increases. By performing nonuniform sensor placement, the optimized placement allocates more sensors to the regions with higher fire ignition probabilities, which increases the fire detection probability therein and maximizes the system utility.

{\bf Annual fire burning area.} Fig.~\ref{fig:compare_area} plots the annual fire burning area. We assume that the fire stops spreading as soon as an event of detecting fires occurs. When the number of sensors is $10^5$, the fire burning areas for biomass uniform placement and optimized placement are $920$ km$^{\text{2}}$ and $478$ km$^{\text{2}}$, respectively, as compared to $10202$ km$^{\text{2}}$ from the CAL FIRE database. Deploying more sensors leads to larger fire detection probability and smaller fire burning area.

{\bf Annual carbon emission.} Fig.~\ref{fig:compare_carbon} plots the annual carbon emission. The amount of carbon emission is given by $F \times 1.2 B_{\rm avg}\times 100$ \cite{sensorbasedIOT}, where $F$ is the fire burning area, $B_{\rm avg}$ is the average above-ground biomass within the fire burning area, and $100$ is a unit-conversion factor. If the fire burning area is inside region $i$, then $B_{\rm avg}=B_i$; if the fire burning area spans several regions, then $B_{\rm avg}$ is the average of $B_i$ of those regions. Both biomass uniform placement and optimized placement implicitly or explicitly consider biomass as a placement rationale which is useful for reducing carbon emission, as biomass is related to both the predictive fire ignition probability (through $p_i^B$) and the actual amount of carbon emission if a fire indeed occurs (fires occurring in regions with small biomass values incur limited carbon emissions). As can be seen, when the number of sensors is $10^5$, the carbon emissions for biomass uniform placement and optimized placement are about $2.5\times10^6$ ton and $1\times10^6$ ton, respectively, which are $4.4$\% and $1.8$\% of the annual carbon emission from the CAL FIRE database.

{\bf Annual savings.} Fig.~\ref{fig:compare_savings} shows the annual monetary savings which are calculated by 
\begin{align} \label{eq:Savings}
{\rm Savings}=\: 
& {\rm Reduction \:of\: carbon\: emissions}\times{\rm Carbon\: price} \nonumber\\
& -{\rm Device\: cost}-{\rm Bandwidth\: cost}
\end{align}
where the reduction of carbon emissions refers to the difference between each sensor placement scheme and CAL FIRE in Fig.~\ref{fig:compare_carbon}. We consider the carbon price of $20\times\gamma$ USD per ton, where $20$ USD is based on the California carbon tax \cite{CarbonTax} and the scaling factor $\gamma$ reflects the savings in terms of fatalities and property damages as a result of fires. We assume $\gamma=10$. We consider two IoT sensor device costs for comparison, i.e., $10$ USD (case A) and $100$ USD (case B) per device. The (worst-case) bandwidth cost was calculated in Sec.~\ref{RA}. As can be seen, significant savings are yielded by both sensor placement schemes (more than $10$ billion USD annually). As the sensor number increases, the costs of devices and bandwidth also increase, leading to diminishing, albeit still sizable, savings. The more drastic decrease in carbon emission when the number of sensors increases from $1\times 10^5$ to $4\times 10^5$ for the biomass uniform placement in Fig.~\ref{fig:compare_carbon} leads to its more noticeable trends of first increasing and then decreasing in the savings, as compared to the optimized placement.

\begin{figure*}[t]
\begin{center}
\subfigure[]{
    \label{fig:bio_colobar_5}
    \includegraphics[width=0.55\columnwidth]{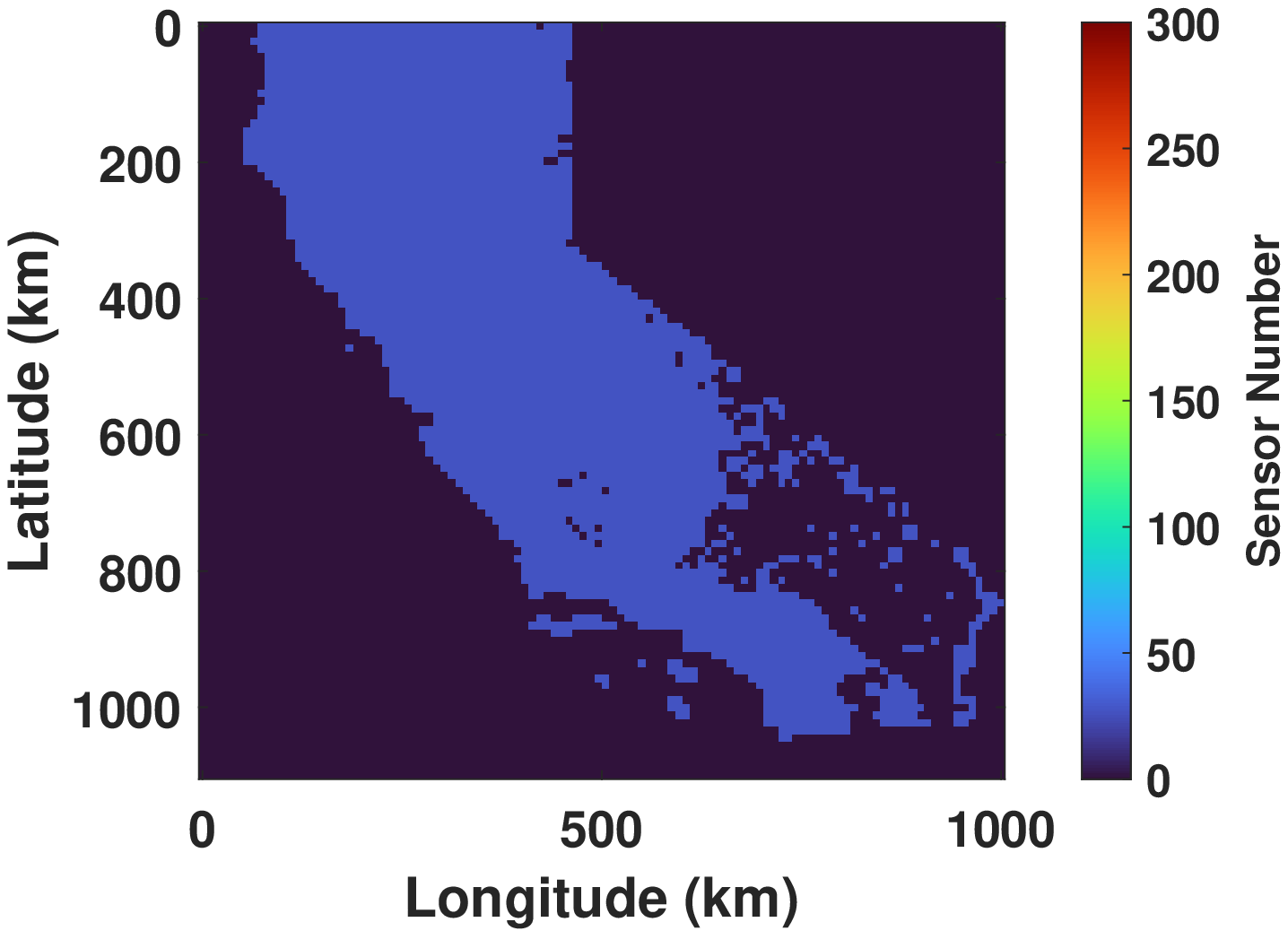}}
\subfigure[]{
    \label{fig:opt_colobar_5}
    \includegraphics[width=0.55\columnwidth]{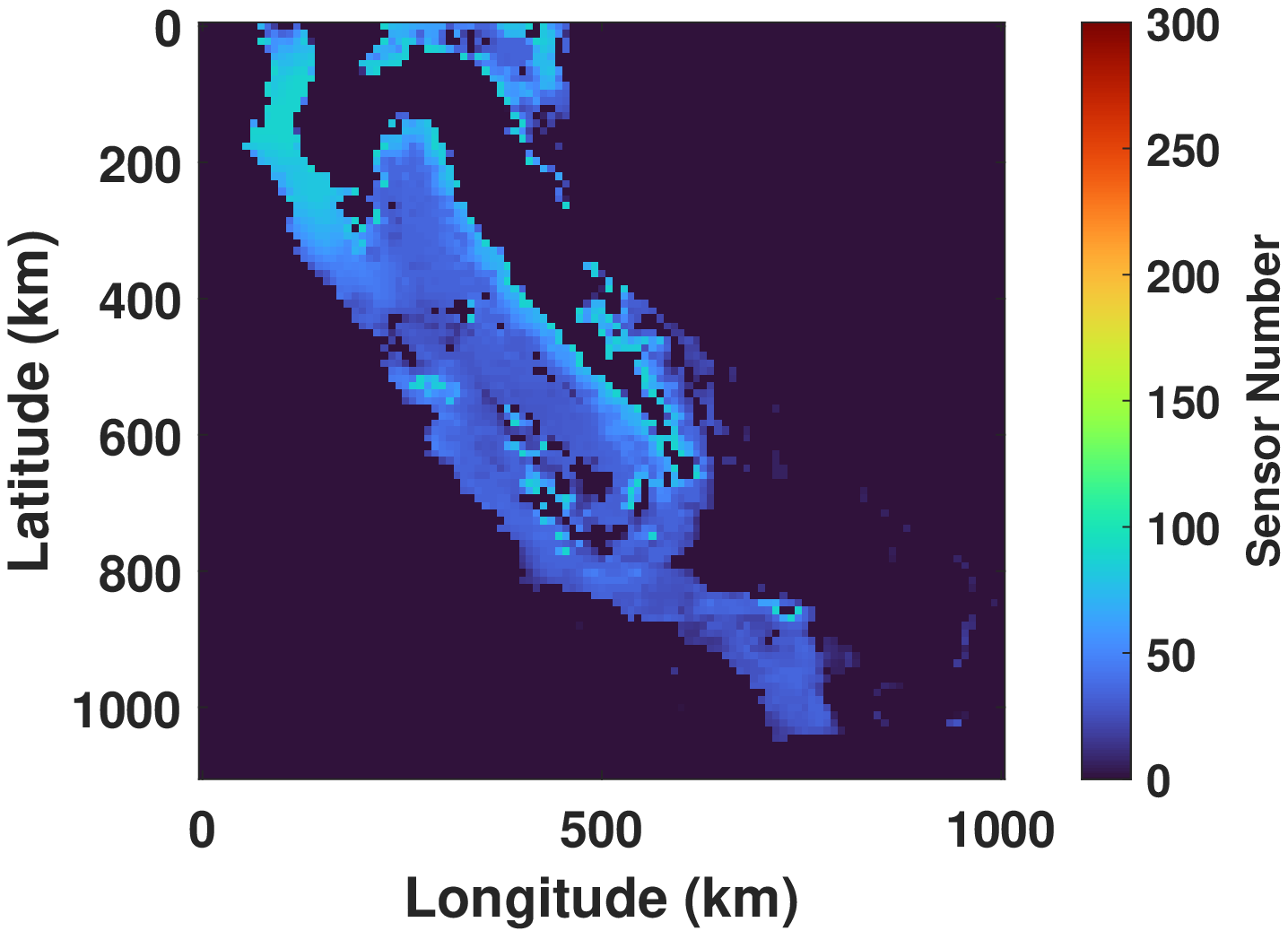}}
\subfigure[]{
    \label{fig:opt8T_colobar_5}
    \includegraphics[width=0.55\columnwidth]{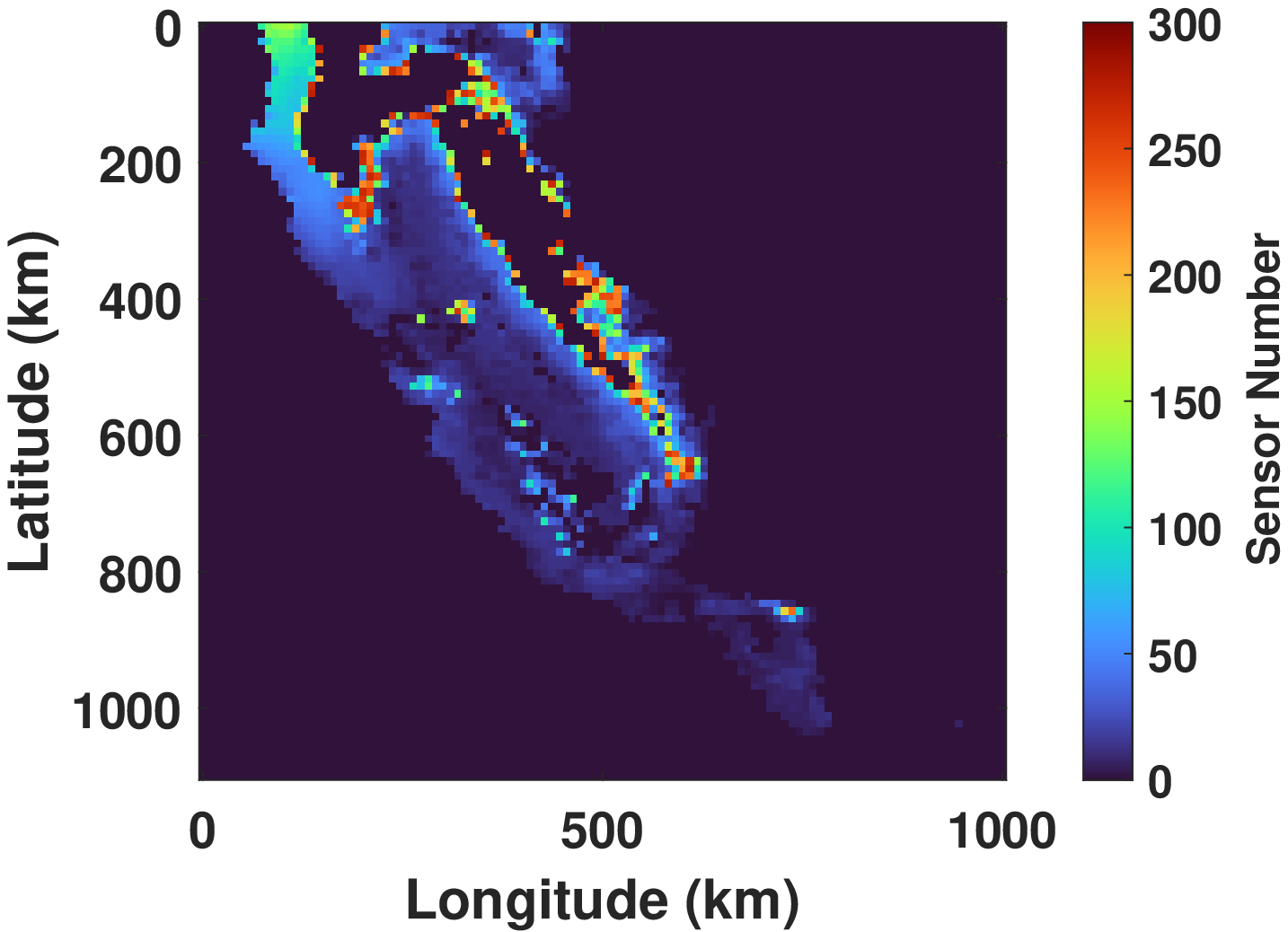}}
\subfigure[]{
    \label{fig:bio_colobar}
    \includegraphics[width=0.55\columnwidth]{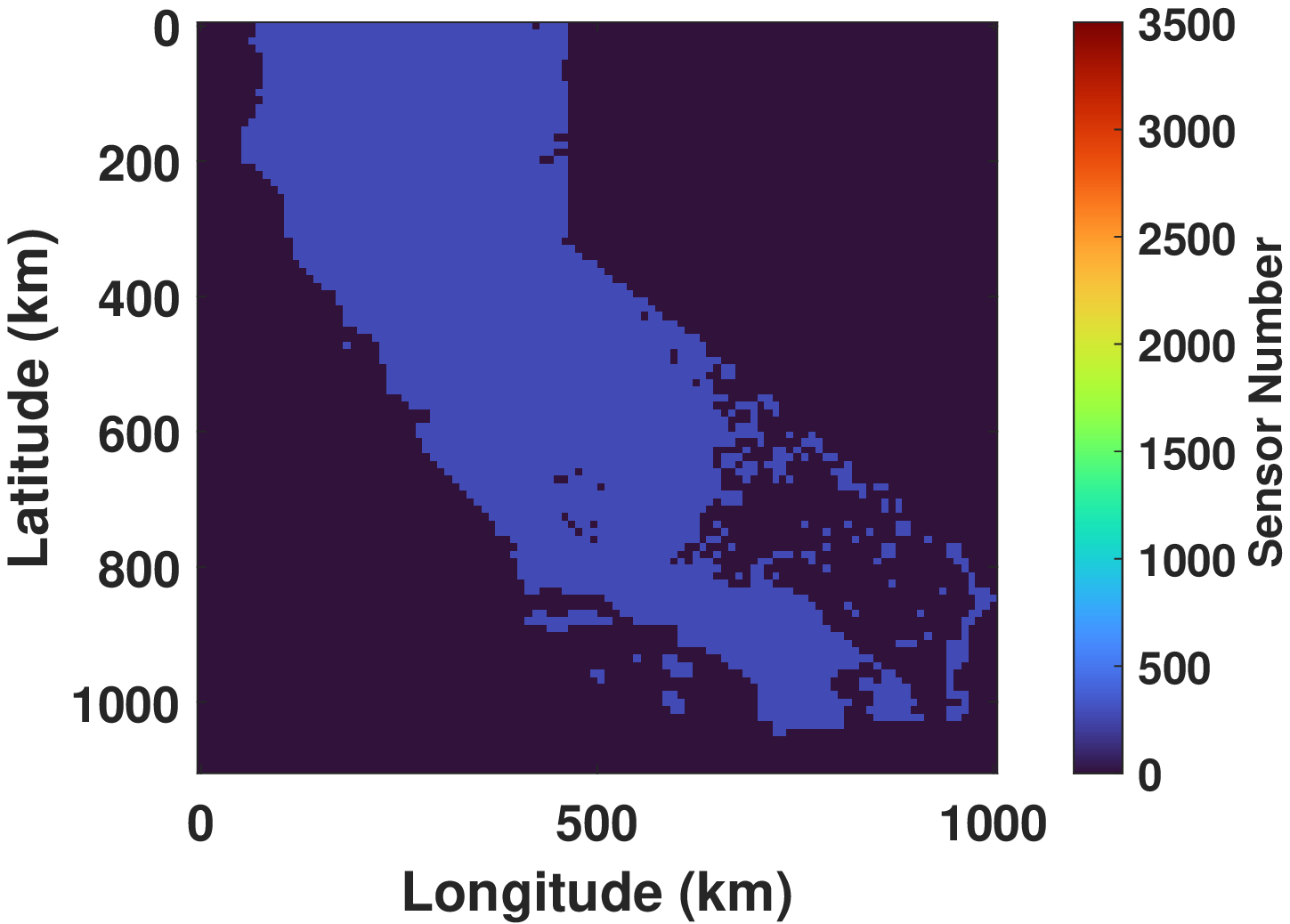}}
\subfigure[]{
    \label{fig:opt_colobar}
    \includegraphics[width=0.55\columnwidth]{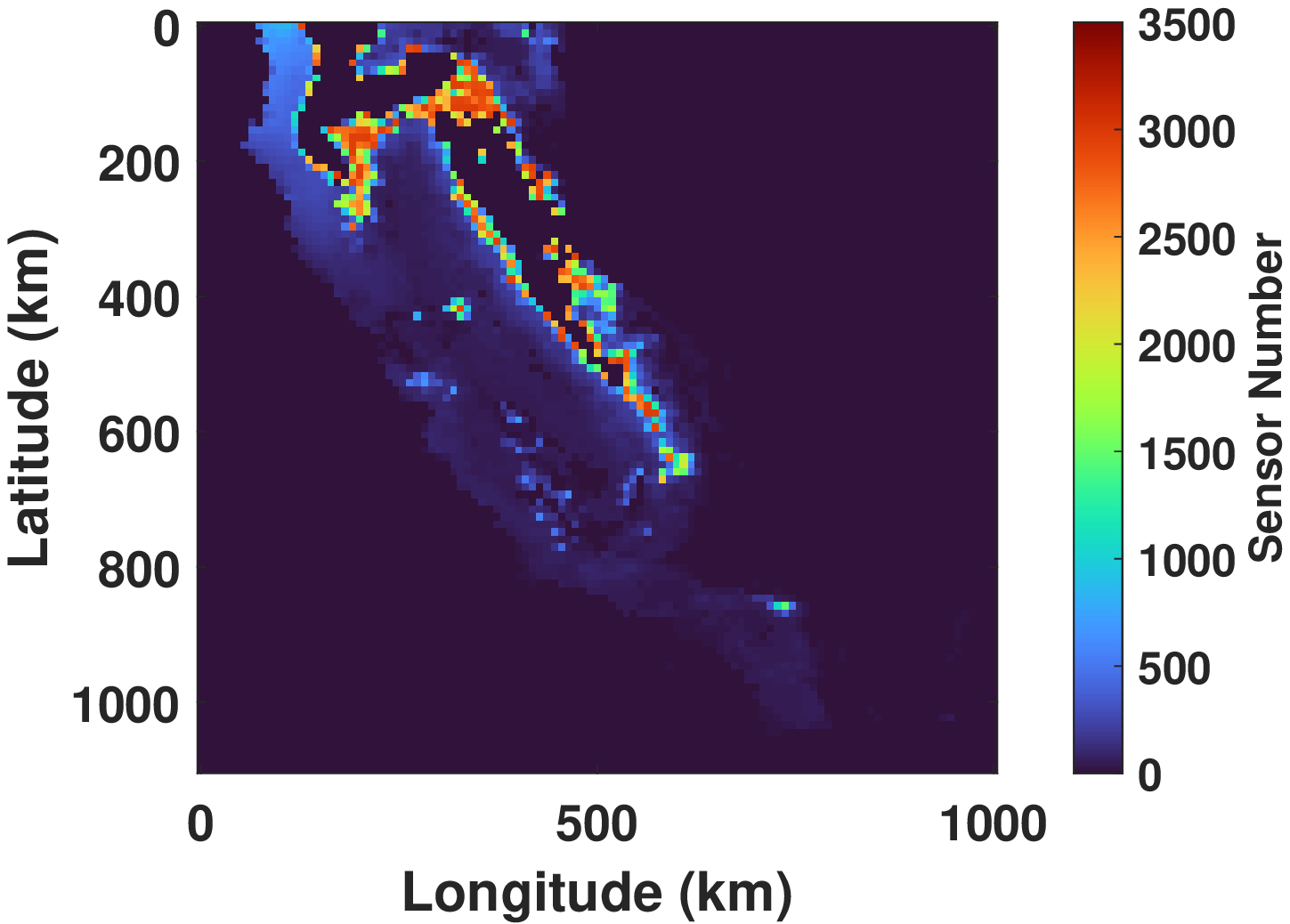}}
\subfigure[]{
    \label{fig:opt8T_colobar}
    \includegraphics[width=0.55\columnwidth]{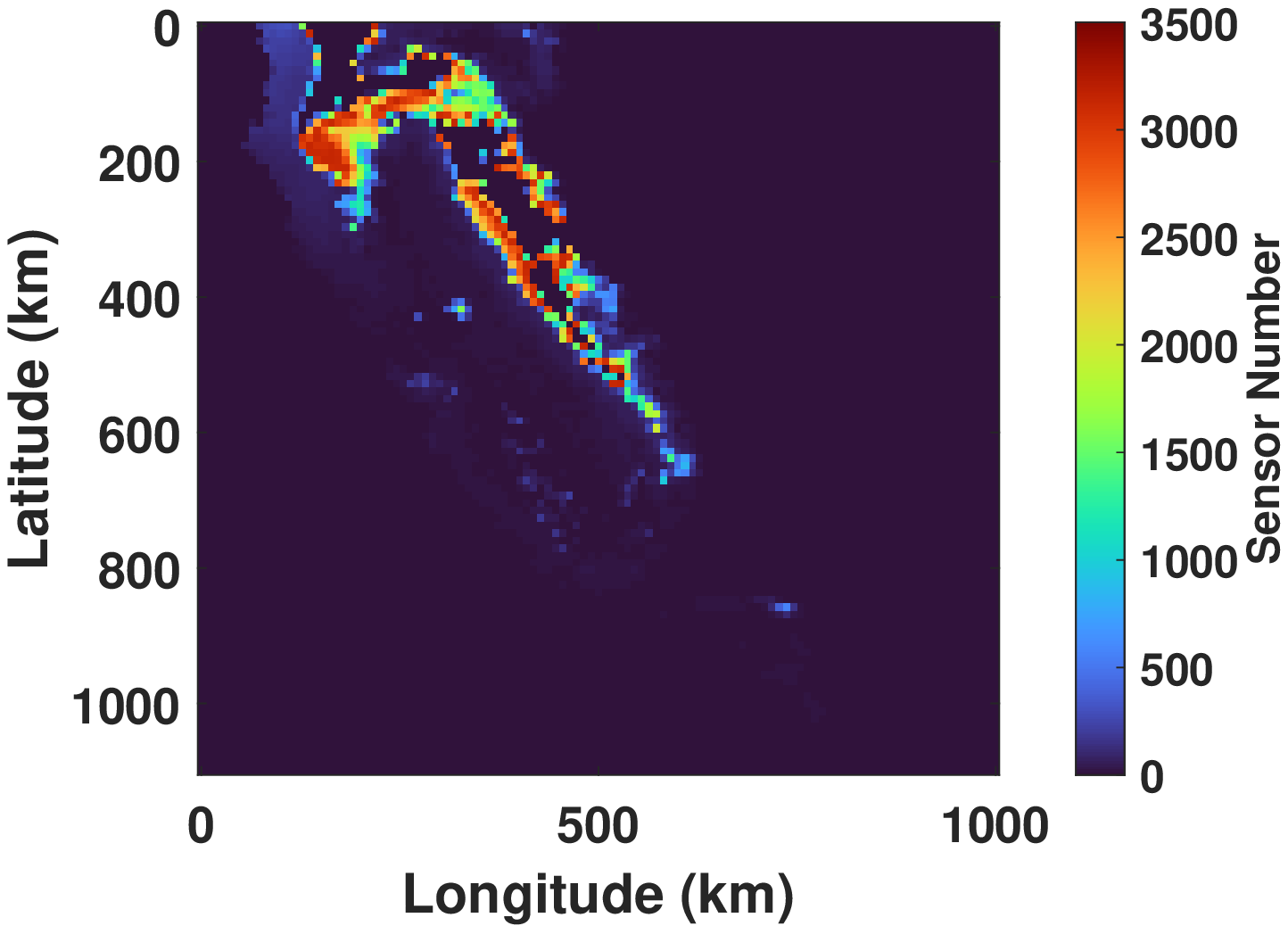}}
\caption{Illustration of sensor placement for a total of $K=10^5$ (first row) and $K=10^6$ (second row) sensors. (a)(d) Biomass uniform placement. (b)(e) Optimized placement with $T=4$. (c)(f) Optimized placement with $T=8$.}
\label{fig:deploy_colorbar}
\end{center}
\vspace{-0.1in}
\end{figure*}

{\bf Illustration of sensor placement.} Fig.~\ref{fig:deploy_colorbar} illustrates the sensor placement and examines the effect of $T$ on the optimized placement. As can be seen, for the biomass uniform placement, sensors are deployed largely uniformly in the geographic area of California, around $28$ and $276$ sensors per region with a positive biomass value when the total number of sensors is $10^5$ and $10^6$, respectively. For the optimized placement, sensors are unevenly deployed. Comparing $T=4$ and $T=8$, a greater $T$ intensifies the effect of the fire spreading rate $u_{p,i}$, increasing $a_i(T)$ and decreasing $A-a_i(T)$ in \eqref{eq:p_FD}, thus resulting in more polarized placement of sensors in different regions in order to increase $p_i^{D}(T)$ and $U(T)$.

\section{Conclusion}

In this paper, we have studied sensor deployment and satellite link budget for sensor-based GEO NTN for wildfire detection. Optimal and biomass-uniform sensor deployment strategies were developed and visualized. Satellite communication budget for a GEO spot beam was analyzed by deriving the uplink SNR for IoT sensors and the worst-case bandwidth required for supporting simultaneous fire event reporting. Simulations using historical California wildfire data demonstrate that the proposed system can significantly reduce fire burned areas and carbon emissions. Specifically, more than $95$\% of carbon reductions and more than $10$ billion USD of savings annually can be yielded when $10^5$--$10^6$ sensors are deployed in California (about $0.24$--$2.4$ sensors per km$^\text{2}$).

\bibliographystyle{IEEEtran}
\bibliography{IEEEabrv,ref}

\end{document}